\documentclass[floats,aps,amssymb,nofootinbib]{revtex4}
\usepackage{amssymb}
\usepackage{graphics}
\usepackage{amsmath}
\usepackage{amsfonts}
\usepackage{bm}
\usepackage[]{latexsym}
\usepackage{float}
\usepackage{graphicx}

\newcommand{\be}{\begin{equation}}\newcommand{\ee}{\end{equation}}
\newcommand{\bea}{\begin{eqnarray}}\newcommand{\eea}{\end{eqnarray}}
\newcommand{\brr}{\begin{array}}\newcommand{\err}{\end{array}}
\newcommand{\bit}{\begin{itemize}}\newcommand{\eit}{\end{itemize}}
\newcommand{\ben}{\begin{enumerate}}\newcommand{\een}{\end{enumerate}}

\newcommand{\ba}{\begin{array}}
\newcommand{\ea}{\end{array}}

\def\lf{\left}

\def\non{\nonumber}\def\pa{\partial}

\def\ri{\right}
\def\al{\alpha}
\def\de{\delta}

\def\si{\sigma}

\def\1{{_{1}}}\def\2{{_{2}}}

\def\noHe0{:\;\!\!\;\!\!:H_e(0):\;\!\!\;\!\!:}
\def\noHm0{:\;\!\!\;\!\!:H_\mu(0):\;\!\!\;\!\!:}

\def\lf{\left}

\def\non{\nonumber}
\def\pa{\partial}

\def\ri{\right}

\def\al{\alpha}
\def\de{\delta}

\def\si{\sigma}

\def\1{{_{1}}}\def\2{{_{2}}}

\begin{document}
\title{Quantum nonlocality in extended theories of gravity}

\author{Victor A. S. V. Bittencourt\footnote{victor.bittencourt@mpl.mpg.de}$^{\hspace{0.3mm}1}$, Massimo Blasone\footnote{blasone@sa.infn.it}$^{\hspace{0.3mm}2,3}$, Fabrizio Illuminati\footnote{filluminati@unisa.it}$^{\hspace{0.3mm}3,4}$, Gaetano Lambiase\footnote{lambiase@sa.infn.it}$^{\hspace{0.3mm}2,3}$, Giuseppe Gaetano Luciano\footnote{gluciano@sa.infn.it}$^{\hspace{0.3mm}2,3}$,  and Luciano Petruzziello\footnote{lupetruzziello@unisa.it}$^{\hspace{0.3mm}3,4}$} \affiliation
{$^1$Max Planck Institute for the Science of Light, Staudtstra\ss e 2, PLZ 91058, Erlangen, Germany.\\
$^2$Dipartimento di Fisica, Universit\`a degli Studi di Salerno, Via Giovanni Paolo II, 132 I-84084 Fisciano (SA), Italy.\\ 
$^3$INFN, Sezione di Napoli, Gruppo collegato di Salerno, Italy.
\\ $^4$Dipartimento di Ingegneria Industriale, Universit\`a degli Studi di Salerno, Via Giovanni Paolo II, 132 I-84084 Fisciano (SA), Italy.}

\date{\today}
\def\be{\begin{equation}}
\def\ee{\end{equation}}
\def\al{\alpha}
\def\bea{\begin{eqnarray}}
\def\eea{\end{eqnarray}}

\begin{abstract}
We investigate how pure-state Einstein-Podolsky-Rosen correlations in the internal degrees of freedom of massive particles are affected by a curved spacetime background described by extended theories of gravity. We consider models for which the corrections to the Einstein-Hilbert action are quadratic in the curvature invariants and we focus on the weak-field limit. We quantify nonlocal quantum correlations by means of the violation of the Clauser-Horne-Shimony-Holt inequality, and show how a curved background suppresses the violation by a leading term due to general relativity and a further contribution due to the corrections to Einstein gravity. Our results can be generalized to massless particles such as photon pairs and can thus be suitably exploited to devise precise experimental tests of extended models of gravity.
\end{abstract}

\vskip -1.0 truecm 

\maketitle

\section{Introduction}\label{I}

The gedanken experiment proposed by Einstein, Podolsky and Rosen (EPR)~\cite{epr} has revealed one of the most striking features of  quantum mechanics (QM): the capability of sharing nonlocal correlations. Originally regarded as a flaw of the theory, such correlations have acquired a more fundamental meaning due to the pioneering work by Bell, whose theorem characterizes the nonlocality of QM in contrast to any local hidden variable (LHV) theory~\cite{bell,BellRev}. Nonlocal correlations cannot be reproduced by any LHV model, thus leading to the violation of a family of inequalities -- the Bell's inequalities. It should be noted that Bell nonlocal correlations and entanglement are in general different types of quantum correlations. While all Bell--nonlocal quantum states are entangled, not all entangled quantum states are Bell nonlocal, a celebrated counterexample being provided by the Werner mixed states~\cite{werner} which are entangled, sharing EPR-type correlations, and yet do not violate any Bell inequality. Indeed, Bell nonlocal (BNL) quantum correlations arising from the violation of Bell inequalities lie at the top of the pyramid of quantum correlations for all quantum states, pure and mixed, and coincide with EPR maximal entanglement for pure states~\cite{EntRev,BellRev}. BNL quantum correlations thus provide one of the fundamental building blocks of quantum information sciences and quantum technologies~\cite{Nielsen}, due both to their intimate hierarchical relation with entanglement~\cite{BellRev,EntRev} and to their own capability as a resource for quantum communication and information tasks~\cite{NLresource}.

Besides its importance for technological applications, the study of BNL quantum correlations may also shed light on the interplay between quantum theory and general relativity. In a seminal work by Terashima and Ueda \cite{ueda} it was observed that a pair of spins initially in a singlet state experiences an apparent degradation of its nonlocal correlations due to gravity-induced spin precession~\cite{ueda}. That is, under a unitary evolution on a curved spacetime, the amount of violation of the Bell's inequality is apparently reduced if one considers a simple rotation of the initial set of operators leading to the maximal violation of Bell's inequality. This occurrence is the result of a succession of infinitesimal local Lorentz transformations describing the evolution of each particle on the curved background, such that the set of operators maximizing the violation of Bell's inequalities depends on the spacetime curvature. The apparent degradation can be understood as the curved spacetime counterpart of the effects of Lorentz boosts on quantum correlations due to Thomas precession~\cite{Flat}. Such phenomenon has also been investigated in connection with Schwarzschild~\cite{ueda,ueda2} and Kerr~\cite{spingrav} metrics.

In the present work, we generalize the above-mentioned studies by analyzing pure-state BNL correlations (that coincide with pure-state EPR correlations) between pairs of spins embedded in a curved spacetime described by \emph{extended} theories of gravity.
For quadratic theories, whose gravitational action is quadratic in the curvature invariants, we show that general relativity (GR) effects such as the ones reported in~\cite{ueda,ueda2,spingrav} can be explicitly separated from those of the extended models. Furthermore, we observe that in the latter it is possible to identify the contribution due to the violation of the equivalence principle~\cite{nordtvedt}, a feature which is typically encountered in extended models of gravity~\cite{capoz}, but which can be found in the GR domain as well when dealing with specific systems such as elementary particles undergoing flavor oscillations and mixing~\cite{wepneutrino}. In particular, for some models which violate the strong equivalence principle (SEP), as for instance $\mathcal{R}^2$ gravity, the degradation of EPR correlations are ascribable to GR terms only. Our results can unravel new physics phenomenology that can be factually tested by virtue of the setup devised in Refs.~\cite{ueda,ueda2,spingrav} and in this paper. The framework is complementary to studies with a focus on gravity implications on wave packets~\cite{spacetimecomm01,spacetimecomm02, spacetimecomm04,spacetimecomm05}. 

The paper is organized as follows: in Sec.~\ref{II} we introduce the vierbein formalism together with the expression for the infinitesimal Wigner rotation in curved spacetime. In Sec.~\ref{III} we briefly summarize the main characteristics involving a peculiar class of extended theories of gravity whose action is quadratic in the curvature invariants. Section~\ref{IV} is devoted to the evaluation of the Wigner angle and Sec.~\ref{V} to the quantification of the EPR correlations, whereas Sec.~\ref{VI} contains several application of our formalism to physically relevant gravitational models. Finally, Sec.~\ref{VII} is reserved for concluding remarks. Throughout the paper, we work with natural units ($c=\hbar=1$) and with the mostly--positive signature convention for the metric, namely $\eta_{ab}=\mathrm{diag}(-,+,+,+)$.

\section{General framework}\label{II}

To properly treat spins in the context of curved spacetime, we first need to introduce the so-called tetrad (or vierbein) formalism~\cite{gravitation}. A tetrad $e^\mu_a$ is defined by
\be\label{tetrad}
g_{\mu\nu}(x)e^\mu_a(x)e^\nu_b(x)=\eta_{ab}\,,
\ee
where $g_{\mu\nu}$ is the metric tensor and $\eta_{ab}$ is the Minkowski metric. In the previous equation, Latin letters denote the ``Lorentzian'' inertial coordinates, whereas Greek letters indicate the ``curved'' general coordinates on the manifold. At each point of the spacetime, vierbein allows us to switch from the general coordinate system to a locally inertial reference frame. A generic tensor defined on the manifold $T^\mu_\nu$ can be converted to its ``flat'' counterpart $T^a_b$ via the vierbein:
\be\label{trans}
T^\mu_\nu\longrightarrow T^a_b=e^a_\mu e^\nu_bT^\mu_\nu\,, 
\ee
with $e^a_\mu$ being the inverse of the vierbein introduced in Eq.~(\ref{tetrad}). Since the above relation holds in all points of the spacetime, we will omit the dependence on $x$ from now on.

The tetrad field is mandatory to define particles with spin $1/2$ on curved backgrounds. A spin $1/2$ particle state is defined as a state belonging to the spin $1/2$ representation of the local Lorentz transformation (LLT) group, and not as a state of the diffeomorphism group~\cite{ueda}. Therefore, if $p^\mu=m u^\mu$ indicates the four-momentum on the manifold normalized as $p^2=-m^2$ (with the four velocity normalized to $u^2=-1$), the momentum associated with a generic spin state in curved spacetime should be given by $p^a=e^a_\mu\,p^\mu$. A generic spin state at the point $x^\mu$ can be then be denoted as $\lf|p^a,\sigma;x\ri\rangle$, with $\sigma$ being the third component of the spin (i.e. $\si=\,\uparrow,\downarrow$). 

Since different local inertial frames can be defined at each point of the curved spacetime, it is licit to analyze the evolution of the state 
after an infinitesimal proper time $d\tau$, during which the particle moves from the point $x^\mu$ to $x'^\mu=x^\mu+u^\mu d\tau$. The momentum change is
\be\label{change}
p^a(x')=p^a(x)+\de p^a(x)\,,
\ee
which is now evaluated in the local inertial frame at $x'^\mu$. Such infinitesimal change can thus be decomposed as
\be\label{change2}
\de p^a=\de p^\mu e^a_\mu+p^\mu\de e^a_\mu\,.
\ee
The first contribution is given in terms of the acceleration $a^\mu=u^\nu\nabla_\nu u^\mu$ due to an external force (i.e. not gravity):
\be\label{term1}
\de p^\mu=ma^\mu d\tau=-\frac{1}{m}\lf(a^\mu p_\nu-p^\mu a_\nu\ri)p^\nu d\tau\,,
\ee
where we used the normalization condition for $p^\mu$ and the identity $p^\mu a_\mu=0$. The second part of Eq.~(\ref{change2}) can be rewritten as
\be\label{term2}
\de e^a_\mu= \chi^a_b\,e^b_\mu d\tau\,,
\ee
with $\chi^a_b$ defined as
\be\label{chi}
\chi^a_b=-u^\mu\omega_{\mu b}^a{}\,, 
\ee
where $\omega_{\mu b}^a=e^a_\nu\nabla_\mu e^\nu_b$ is the connection one-form~\cite{gravitation}. If we now plug Eqs.~(\ref{term1}) and (\ref{term2}) into~(\ref{change2}) we get
\be\label{lorentz}
\de p^a=\lambda^a_b\,p^bd\tau\,,
\ee
where
\be\label{lorentz2}
\lambda^a_b=-\frac{1}{m}\lf(a^ap_b-p^aa_b\ri)+\chi^a_b\,
\ee
is an infinitesimal LLT. As a matter of fact, as the particle moves in the infinitesimal proper time interval $d\tau$, the momentum in the local inertial frame transforms as 
\be\label{lorentz3}
\Lambda^a_b=\delta^a_b+\lambda^a_bd\tau\,.
\ee
Correspondingly, a spin state should transform via a representation of such LLT transformation.

As shown by Wigner~\cite{wig}, Lorentz transformations act as unitary operators on quantum states. The specific form of the unitary transformation associated with a given frame transformation depends on the group representation to which the state belongs. The simplest example is the action of a proper Lorentz transformation connecting two inertial frames on the state of a particle carrying spin. The latter is deemed as a suitable representation of the Lorentz (or Poincar\'{e}) group~\cite{WuTung}. 

For a particle state with momentum $p$ and spin $s$, a Lorentz transformation $\Lambda$ connecting two inertial frames $O$ and $O^\prime$, acts on the particle state as a unitary operator $U$
\begin{equation}
\vert \Lambda p,\si^\prime ; s \rangle= U\lf(\Lambda\ri) \vert p, \si; s \rangle,
\end{equation}
were $p$ ($\Lambda p = p^\prime$) is the momentum of the particle in the frame $O$ ($O^\prime$). To obtain the operator $U\lf(\Lambda\ri)$, we consider the transformation from a standard reference frame to the particle's rest frame $R$. This is the standard procedure outlined, for instance, in \cite{WuTung,weinberg}. Since the transformation $R \rightarrow O \rightarrow O^\prime$ is in general a succession of two non-collinear boosts, it can be decomposed as a pure boost and a rotation, i.e. the Wigner rotation. This allows the description of frame transformations on spin states via a suitable representation of a Wigner rotation acting on the spin. If the boost describing the transformation $O \rightarrow O^\prime$ is collinear with the momentum $p$, no rotation is induced.

The procedure is formalized as follows. The momentum $p$ can be written via a boost $L$ acting on the rest-frame momentum $k$: $p= L(p) k$, where $L(p)$ is a Lorentz boost with elements
\be\label{standard}
L^0_0=\xi\,, \quad L^i_0=L^0_i=\frac{p^i}{m}\,, \quad L^i_k=\de_{ik}+\lf(\xi-1\ri)\frac{p^ip^k}{\lf|\vec{p}\ri|^2}
\ee
and $\xi=\sqrt{\lf|\vec{p}\ri|^2+m^2}$, with the indexes $i,k=1,2,3$. We can thus write
\begin{equation}
U\lf(\Lambda\ri) \vert p, \si; s \rangle = U\lf(\Lambda L(p) \ri) \vert k, \si; s \rangle = U\lf( L(\Lambda p) \ri) U\lf(L^{-1}(\Lambda p) \Lambda L(p) \ri) \vert k, \si; s \rangle.
\end{equation}
The operator $U\lf( L(\Lambda p) \ri)$ represents the frame transformation  $R \rightarrow O^\prime$ via a boost from the rest frame to the frame in which the particle has momentum $\Lambda p$. The second operator on the left hand side, $U\lf(L^{-1}(\Lambda p) \Lambda L(p) \ri)$, represents the transformation $R \rightarrow O \rightarrow O^\prime \rightarrow R$, in which the momentum is transformed as $k \rightarrow p \rightarrow \Lambda p \rightarrow k$. The overall result of the latter is the transformation
\begin{equation}
\label{wigner}
W (\Lambda, p) =L^{-1}(\Lambda p) \Lambda L(p)
\end{equation}
called a Wigner rotation. The set of all $W$ forms a subgroup of the proper orthochronous Lorentz group, called the Wigner's little group. The action of a Lorentz transformation on the quantum state is thus
\begin{eqnarray}
U\lf(\Lambda\ri) \vert p, \si; s \rangle &=& U \lf( W (\Lambda, p) \ri) U\lf( L(\Lambda p) \ri) \vert k, \si; s \rangle  \nonumber \\
&=& U \lf( W (\Lambda, p) \ri) \vert \Lambda p,\si; s \rangle = \displaystyle \sum_{\si^\prime} D_{\si,\si^\prime} ^{(s)} \left( W (\Lambda, p) \right) \vert \Lambda p, \si^\prime; s \rangle,
\end{eqnarray}
where $U \lf( W (\Lambda, p) \ri)$ is a representation of the Wigner's little group corresponding to the spin $s$ with elements $D_{\si,\si^\prime} ^{(s)} \left( W (\Lambda, p) \right) $. The spin of the particle is then rotated by an amount that depends on the momentum of the particle. It should be noted that the same formalism can also be applied to the description of particle states via Dirac bispinors~\cite{bispinors}. Wigner rotations on quantum states find fruitful applications in relativistic quantum mechanics, in particular for the study of the behavior of information and quantum correlations under frame transformations.

Now, we turn our attention to spin $1/2$ particles; consequently, we have~\cite{ueda,weinberg}
\be\label{spintran}
U\lf(\Lambda\ri)\lf|p^a,\si;x\ri\rangle=\sum_{\si'}D_{\si'\si}^{(1/2)}\lf(W\ri)\lf|\Lambda p^a,\si';x\ri\rangle\,,
\ee
with $D$ being a unitary $2\times2$ matrix that acts on the spin of a particle according to the Wigner rotation $W$. For massive particles, the little group is the usual rotation group in 3 dimensions, i.e. $SO(3)$, and accordingly $D^{(1/2)}$ belongs to $SU(2)$, thereby taking the form $\rm{exp}[- i (\phi /2) {\bm{\sigma}} \cdot {\bold{n}}]$, with rotation angle $\phi$ and direction $\bm{n}$ defined by the specific details of the problem at hand.

Although the above ideas were presented in the context of proper Lorentz transformations in flat spacetime, frame transformations in curved spacetime can be conceived as successions of infinitesimal Lorentz transformations whose effects on spin states can be obtained via a suitable integration of the induced rotations. By recalling Eq.~(\ref{lorentz3}), one can thus check~\cite{ueda} that the infinitesimal expression for the Wigner rotation~(\ref{wigner}) is $W^a_b=\de^a_b+\vartheta^a_bd\tau$, with
\be\label{wigner2}
\vartheta^i_k=\lambda^i_k+\frac{\lambda^i_0p_k-\lambda_{k0}p^i}{p^0+m}\,,
\ee
whereas all other terms are zero.

\section{Quadratic theories of gravity}\label{III}

Before moving forward, we briefly recall the main aspects of the modified theories of gravity we are going to analyze. The starting point is the most general action that is torsion-free, parity invariant and quadratic in the curvature invariants, given by~\cite{Asorey:1996hz,Modesto:2011kw,Biswas:2011ar,Biswas:2013cha,Biswas:2016etb}
\begin{equation}
S=\frac{1}{2\kappa^2}\int d^4x\sqrt{-g}\left\lbrace \mathcal{R}+\frac{1}{2}\Bigl[\mathcal{R}\mathcal{F}_1(\Box)\mathcal{R}+\mathcal{R}_{\mu\nu}\mathcal{F}_2(\Box)\mathcal{R}^{\mu\nu}+\mathcal{R}_{\mu\nu\rho\sigma}\mathcal{F}_3(\Box)\mathcal{R}^{\mu\nu\rho\sigma}\Bigr]\right\rbrace,
\label{quad-action}
\end{equation}
with $\kappa=\sqrt{8\pi G}$, $\Box=g^{\mu\nu}\nabla_{\mu}\nabla_{\nu}$ being the d'Alembert operator in curved spacetime and $\mathcal{F}_i(\Box)$ are operators of $\Box$ that can be either local or nonlocal
\begin{equation}
\mathcal{F}_i(\Box)=\sum\limits_{n=0}^{N}f_{i,n}\Box^n\,, \qquad i=1,2,3.
\end{equation}
%
Since our aim is to treat EPR correlations in the presence of a weak gravitational field, we expand the action~\eqref{quad-action} in small fluctuations around the Minkowski metric
\begin{equation}
g_{\mu\nu}=\eta_{\mu\nu}+\kappa h_{\mu\nu}\,.\label{lin-metric}
\end{equation}
It can then be shown that $\mathcal{R}_{\mu\nu\rho\sigma}\mathcal{F}_3(\Box)\mathcal{R}^{\mu\nu\rho\sigma}$ in Eq.~\eqref{quad-action} can be rewritten as a function of the other two terms that depend on the scalar curvature and the Ricci tensor~\cite{ours}. 

By accounting for the previous observation and substituting Eq.~\eqref{lin-metric} into~\eqref{quad-action} we obtain~\cite{Biswas:2011ar}
\begin{equation}
S=\frac{1}{4}\int d^4x\left[\frac{1}{2}h_{\mu\nu}a(\Box)\Box h^{\mu\nu}-h_{\mu}^{\sigma}a(\Box)\partial_{\sigma}\partial_{\nu}h^{\mu\nu}+hc(\Box)\partial_{\mu}\partial_{\nu}h^{\mu\nu}-\frac{1}{2}hc(\Box)\Box h
+\frac{1}{2}h^{\lambda\sigma}\frac{a(\Box)-c(\Box)}{\Box}\partial_{\lambda}\partial_{\sigma}\partial_{\mu}\partial_{\nu}h^{\mu\nu}\right] ,
\label{lin-quad-action}
\end{equation}
where $h\equiv\eta_{\mu\nu}h^{\mu\nu}$ and
\begin{equation}
a(\Box)=1+\frac{1}{2}\mathcal{F}_2(\Box)\Box\,, \qquad
c(\Box)=1-2\mathcal{F}_1(\Box)\Box-\frac{1}{2}\mathcal{F}_2(\Box)\Box\,.
\end{equation}
The ensuing field equations then read
\begin{equation}
a(\Box)\left(\Box h_{\mu\nu}-\partial_{\sigma}\partial_{\nu}h_{\mu}^{\sigma}-\partial_{\sigma}\partial_{\mu}h_{\nu}^{\sigma}\right)+c(\Box)\left(\eta_{\mu\nu}\partial_{\rho}\partial_{\sigma}h^{\rho\sigma}+\partial_{\mu}\partial_{\nu}h-\eta_{\mu\nu}\Box h\right)+ \frac{a(\Box)-c(\Box)}{\Box}\partial_{\mu}\partial_{\nu}\partial_{\rho}\partial_{\sigma}h^{\rho\sigma}=-2\kappa^2 T_{\mu\nu},
\label{lin-field-eq}
\end{equation}
where 
\begin{equation}
T_{\mu\nu}=-\frac{2}{\sqrt{-g}}\frac{\delta S_m}{\delta g^{\mu\nu}}
\end{equation}
is the stress-energy tensor which is the source of gravity and $S_m$ is the matter action.

The line element in the presence of a static point-like source of mass $M$ in isotropic coordinates can then be cast in the form
\begin{equation}
ds^2=-(1+2\phi)dt^2+(1-2\psi)(dr^2+r^2d\Omega^2)\,,\label{isotr-metric}
\end{equation}
where $d\Omega^2=d\theta^2+\sin^2\theta\,d\varphi^2$, while $\phi$ and $\psi$ are the metric potentials stemming from the chosen setting for the stress-energy tensor, namely
\begin{equation}
T_{\mu\nu}=M\,\delta_{\mu}^t\,\delta_{\nu}^t\,\delta(\vec{r})\,.
\end{equation}
Indeed, by noting that $\kappa h_{00}=-2\phi$, $\kappa h_{ij}=-2\psi\delta_{ij}$, $\kappa h=2(\phi-3\psi)$, $T=\eta_{\rho\sigma}T^{\rho\sigma}\simeq-T_{00}=-\rho$ and recalling that the source is static (i.e., $\Box\simeq \nabla^2$),  the field equations for the two metric potentials are
\begin{equation}
\frac{a(a-3c)}{a-2c}\nabla^2\phi(r)=8\pi GM\delta(\vec{r})\,, \qquad
\frac{a(a-3c)}{c}\nabla^2\psi(r)=- 8\pi GM\delta(\vec{r})\,,
\label{field-eq-pot}
\end{equation}
the solutions of which are given by
\begin{eqnarray}\non
\phi(r)&=&-8\pi Gm\int \frac{d^3k}{(2\pi)^3}\frac{1}{k^2}\frac{a-2c}{a(a-3c)}e^{i\vec{k}\cdot \vec{r}}=-\frac{4Gm}{\pi r}\int_0^{\infty}dk\frac{a-2c}{a(a-3c)}\frac{{\rm sin}(kr)}{k}\,,\\[2mm]
\psi(r)&=&8\pi Gm\int \frac{d^3k}{(2\pi)^3}\frac{1}{k^2}\frac{c}{a(a-3c)}e^{i\vec{k}\cdot \vec{r}}=\frac{4Gm}{\pi r}\int_0^{\infty}dk\frac{c}{a(a-3c)}\frac{{\rm sin}(kr)}{k}\,,
\label{fourier-pot}
\end{eqnarray}
where now $a$ and $c$ are functions of $k^2$. Clearly, for $a=c=1$, we recover the standard GR result. At this point, we have everything we need to perform our considerations.

\section{Wigner rotation angle for quadratic theories of gravity}\label{IV}

To compute the effects of curved spacetime as described in the previous Section in the context of EPR correlations, we have to evaluate all the tetrads and the corresponding Wigner rotation angle introduced in Sec.~\ref{II} starting from the line element defined in Eq.~\eqref{isotr-metric}. As already anticipated, we work in the weak-field approximation, which allows us to neglect higher-order terms in gravitational potentials $\phi$ and $\psi$ and their derivatives throughout our calculations.
Hence, we start by computing the vierbein fields associated to the metric~\eqref{isotr-metric}, which are given by
\be\label{lintetrad}
e^t_0=1-\phi\,, \quad e^r_1=1+\psi\,, \quad e^\theta_2=\frac{1+\psi}{r}\,, \quad e^\varphi_3=\frac{1+\psi}{r\sin\theta}\,,
\ee
together with their inverse
\be\label{invtetrad}
e^0_t=1+\phi\,, \quad e^1_r=1-\psi\,, \quad e^2_\theta=r(1-\psi)\,, \quad e^3_\varphi=r\sin\theta(1-\psi)\,.
\ee
The non-vanishing connection one-forms are then evaluated as
\bea\non
\omega^0_{t1}=\omega^1_{t0}=\pa_r\phi\,, \qquad \omega^2_{\theta1}=-\omega^1_{\theta2}=1-r\pa_r\psi\,, \qquad \omega^3_{\theta3}=\cot\theta\,,\\[2mm]\label{oneform}
\omega^3_{\varphi1}=-\omega^1_{\varphi3}=\sin\theta\lf(1-r\pa_r\psi\ri)\,, \qquad \omega^3_{\varphi2}=-\omega^2_{\varphi3}=\cos\theta\,, \qquad
\eea
Without loss of generality, we assume that the physical setup lies on the equatorial plane $\theta=\pi/2$ and is at rest (i.e. locally inertial) in the curved spacetime. Furthermore, we suppose that the EPR source is located at $\varphi=0$ whereas the observers occupy the positions denoted with $\pm\varphi$ (see Fig.~\ref{figure} below). 
\begin{figure}[ht!]
  \centering
    \includegraphics[width=9.5cm]{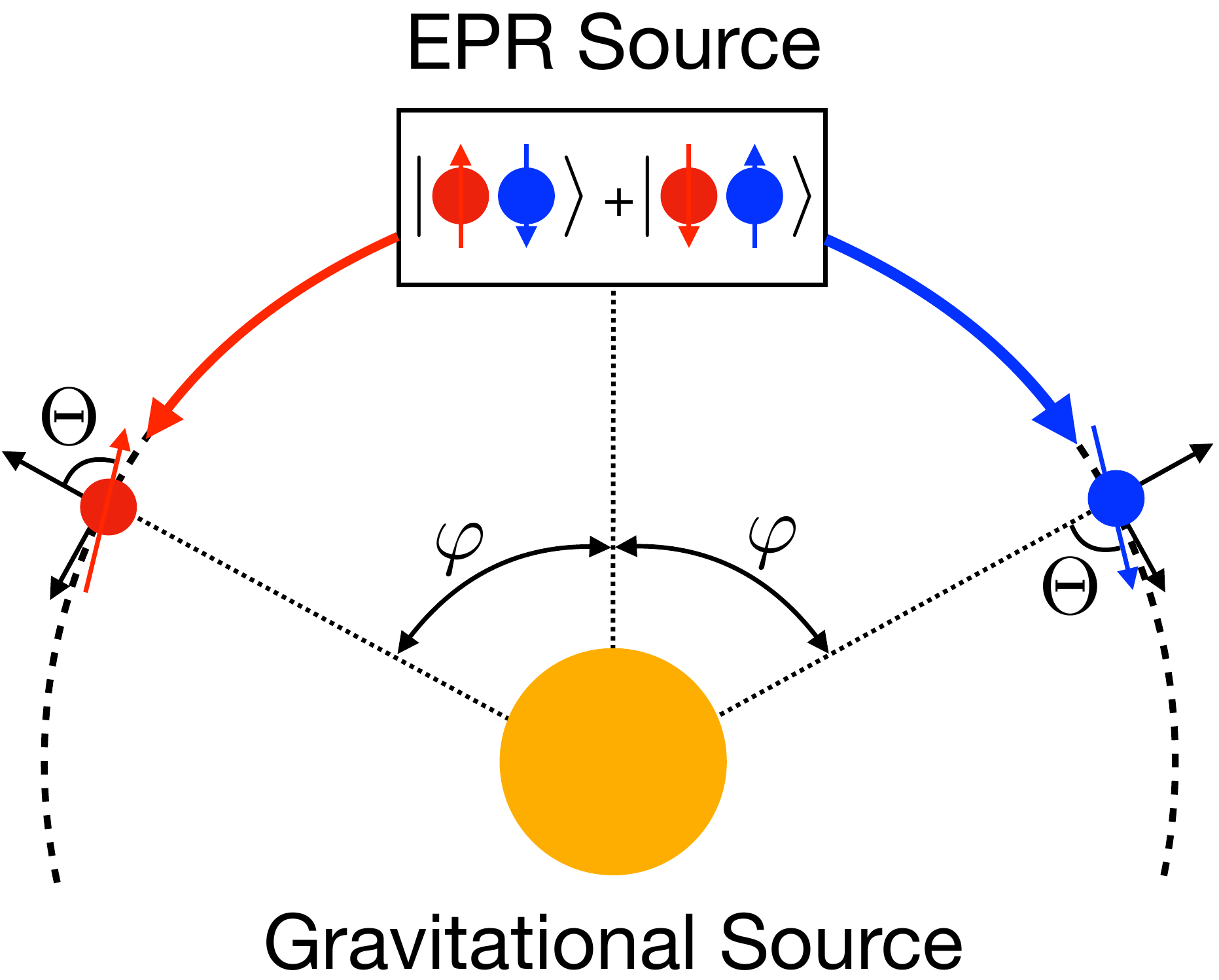}
    \caption{A pair of spin-$1/2$ particles in a singlet state is produced at $\varphi=0$. Each particle propagates in opposite directions on a circular orb around a gravitational source. After the evolution, the spins will be rotated due to the spacetime curvature.}
  \label{figure}
\end{figure}
In this scenario, the four-velocity is given by the compact expression
\be\label{fourvel}
u^\mu(x)=\lf((1-\phi)\cosh\zeta,0,0,\frac{1+\psi}{r}\sinh\zeta\ri)\,,
\ee
where $\zeta$ is the rapidity in the local inertial frame, and the ``local''four-velocity takes the standard form
\be\label{locvel}
u^a=\lf(\cosh\zeta,0,0,\sinh\zeta\ri)\,.
\ee
Since this is not a geodesic trajectory, an external force must act to balance the gravitational interaction. The acceleration produced by such force has only the radial component given by
\be\label{acc}
a^r=\pa_r\phi\cosh^2\zeta-\frac{1+2\psi-r\pa_r\psi}{r}\sinh^2\zeta\,.
\ee
The infinitesimal Wigner rotation~\eqref{wigner2} can now be computed by first evaluating $\chi^a_b$ defined in Eq.~\eqref{chi}, namely
\be\label{newchi}
\chi^1_0=\chi^0_1=-\pa_r\phi\cosh\zeta\,, \qquad \chi^3_1=-\chi^1_3=-\frac{1+\psi-r\pa_r\psi}{r}\sinh\zeta\,,
\ee
which gives the following LLT
\be\label{illy}
\lambda^0_1=\lambda^1_0=\cosh\zeta\sinh^2\zeta\lf(\pa_r\phi-\frac{1+\psi-r\pa_r\psi}{r}\ri)\,, \qquad \lambda^1_3=-\lambda^3_1=\cosh^2\zeta\sinh\zeta\lf(\pa_r\phi-\frac{1+\psi-r\pa_r\psi}{r}\ri)\,.
\ee
The only non-vanishing $\vartheta^i_k$ appearing in Eq.~\eqref{wigner2} are
\be\label{infwig}
\vartheta^3_1=-\vartheta^1_3=\cosh\zeta\sinh\zeta\lf(\pa_r\phi-\frac{1+\psi-r\pa_r\psi}{r}\ri)\,.
\ee
Finally, the exact Wigner transformation over a finite proper time interval is given by iterating the infinitesimal transformation. This process turns out to build up a Dyson series, whose solution is well-known and converge to~\cite{ueda}
\be\label{finitewigner}   
W^3_1=T\,\exp\lf[{\int_{\tau_i}^{\tau_f}\vartheta^3_1(x(\tau'))d\tau'}\ri]=\exp\lf[{\vartheta^3_1\lf(\tau_f-\tau_i\ri)}\ri]\,, \qquad 
\ee
where in the second step we have made use of the fact that $\vartheta^3_1$ is constant during the motion.

\section{Nonlocality and EPR correlations}\label{V}

Let us now consider an EPR source emitting a pair of entangled particles, labeled A and B, in opposite directions with constant four-momenta $p^a_\pm=(m\cosh\zeta,0,0,\pm m\sinh\zeta)$ in the spin-singlet pure state
\be\label{instate}
\vert \psi \rangle = \frac{1}{\sqrt{2}}\lf(|p^a_+,\uparrow;0\rangle_A |p^a_-,\downarrow;0\rangle_B-|p^a_+,\downarrow;0\rangle_A |p^a_-,\uparrow;0\rangle_B \ri)\,,
\ee
in which we have emphasized only the coordinate $\varphi$ for the sake of brevity. This state encodes nonlocal correlations between the spins degrees of freedom, which we now investigate in terms of a suitable Bell's inequality~\cite{BellRev}.  As established by Bell's theorem, there exist nonlocal quantum correlations that cannot be accounted for by any set of LHV models; this tenet ultimately leads to the Bell's inequalities.


The simplest form of Bell's inequality is the Clauser, Horne, Shimony and Holt (CHSH) inequality \cite{CHSH} that has been used for studying nonlocality in relativistic scenarios \cite{Flat}. The CHSH inequality provides a tool for testing the LHV model of a given correlation between two dichotomous variables, which in the present case we identify with the two spins. At this point, consider that two sets of measurements $\{ \hat{A}_1, \hat{A}_2 \}$ and $\{ \hat{B}_1, \hat{B}_2 \}$  are performed on the spin of particles A and B, respectively. If the correlations shared between the spins in a given pure state $\vert \Psi \rangle$ are local (and therefore reproducible by a suitable LHV theory), then for all sets of measurements
\begin{equation}
\label{CHSHQuant}
\mathcal{S}[\vert \Psi \rangle] = \vert \langle \hat{A}_1 \hat{B}_1 \rangle + \langle \hat{A}_1 \hat{B}_2 \rangle +\langle \hat{A}_2 \hat{B}_1 \rangle -\langle \hat{A}_2 \hat{B}_2 \rangle \vert \le 2\,,
\end{equation}
where $\langle \hat{A}_i \hat{B}_j \rangle = \langle \Psi  \vert \hat{A}_i \hat{B}_j \vert \Psi  \rangle $. If for some set of observables the above inequality is violated, it means that the correlations shared between the spins are nonlocal. It should be pointed out one more time that, in general, non-local correlations are not equivalent to entanglement, although the concepts possess an intrinsic relation~\cite{EntRev}.

The pure state~\eqref{instate} realizes a canonical example of violation of the CHSH inequality. Indeed, the sets 
\begin{equation}
\label{Observables}
\begin{aligned}
\hat{A}_1 =\hat{S}_x^{(A)}&, \quad \hat{A}_2=\hat{S}_y^{(A)}, \\[2mm]
\hat{B}_1 = -\frac{\hat{S}_x^{(B)} + \hat{S}_y^{(B)}}{\sqrt{2}}&, \quad \hat{B}_2 =\frac{\left (-\hat{S}_x^{(B)} + \hat{S}_y^{(B)} \right)}{\sqrt{2}}
\end{aligned}
\end{equation}
lead to the maximum violation of the inequality $\mathcal{S}[\vert \psi \rangle] = 2 \sqrt{2}$. The evolution of~\eqref{instate} in a curved spacetime changes the degree of violation of the CHSH inequality, which we can evaluate by computing the transformation describing the evolution of the state in terms of Wigner rotations.

As the particles A and B reach the respective observer after a finite proper time $\tau={r\varphi}(1-\psi)/\sinh\zeta$, the Wigner transformation is nothing but a rotation about the 2-axis~\cite{ueda,spingrav} given as (in its matrix representation)
\be\label{wignermatrix}
\mathbb{W}\lf(\pm\varphi\ri)=\begin{pmatrix} 1 & 0 & 0 & 0 \\ 0 & \cos\Theta & 0 & \pm\sin\Theta \\ 0 & 0 & 1 & 0 \\ 0 & \mp\sin\Theta & 0 & \cos\Theta \end{pmatrix}\,.
\ee
From Eq.~\eqref{finitewigner}, we can extract the value for $\Theta$, that is
\be\label{btheta}
\Theta=\frac{r\varphi(1-\psi)}{\sinh\zeta}\vartheta^1_3=\varphi\cosh\zeta\bigl[1-r\pa_r\lf(\phi+\psi\ri)\bigr]\, .
\ee
This evolution and the whole physical setup are reported schematically in Fig.~\ref{figure}.

The transformation describing the changes in the spin state~\eqref{spintran} associated to Eq.~\eqref{wignermatrix} is represented by~\cite{ueda,spingrav} 
\be\label{spinhalf}
D_{\si'\si}^{(1/2)}=e^{\mp i\frac{\si_y}{2}\Theta}\,,
\ee
where $\si_y$ is the Pauli matrix. This expression can also be deduced from the general infinitesimal $\vartheta^a_b$ 
\be\label{infspin}
D_{\si'\si}^{(1/2)}=\textbf{1}+\frac{i}{2}\lf(\vartheta_{23}\si_x+\vartheta_{31}\si_y+\vartheta_{12}\si_z\ri)d\tau
\ee
specialized to our case, which contemplates only the second term as non-vanishing.

However, since during the evolution the spin-singlet state gets linearly superposed with the spin-triplet state, measurements of the spin along the same direction are in general not anticorrelated in the local reference frame at $\pm\varphi$~\cite{ueda,spingrav}. To get rid of this undesired influence, we should simply perform a rotation of the bases about the 2-axis by $\mp\varphi$ in the point indicated by $\pm\varphi$. In other words, we need to work with the following rotated states:
\bea\non
|p^a_\pm,\uparrow;\pm\varphi\rangle'&=&\cos\frac{\varphi}{2}|p^a_\pm,\uparrow;\pm\varphi\rangle\pm\sin\frac{\varphi}{2}|p^a_\pm,\downarrow;\pm\varphi\rangle\,,\\[2mm]\label{newstate}
|p^a_\pm,\downarrow;\pm\varphi\rangle'&=&\mp\sin\frac{\varphi}{2}|p^a_\pm,\uparrow;\pm\varphi\rangle+\cos\frac{\varphi}{2}|p^a_\pm,\downarrow;\pm\varphi\rangle\,.
\eea
The evolved state can thus be rewritten as 
\be\label{finstate}
\vert \psi^\prime \rangle = \frac{1}{\sqrt{2}}\Bigl[\cos\Delta\Bigl(|p^a_+,\uparrow;\varphi\rangle'|p^a_-,\downarrow;-\varphi\rangle'-|p^a_+,\downarrow;\varphi\rangle'|p^a_-,\uparrow;-\varphi\rangle'\Bigr)+\sin\Delta\Bigl(|p^a_+,\uparrow;\varphi\rangle'|p^a_-,\uparrow;-\varphi\rangle'+|p^a_+,\downarrow;\varphi\rangle'|p^a_-,\downarrow;-\varphi\rangle'\Bigr)\Bigr]\,,
\ee
where now
\be\label{delta}
\Delta=\Theta-\varphi=\varphi\,\Bigl\{\cosh\zeta\bigl[1-r\pa_r\lf(\phi+\psi\ri)\bigr]-1\Bigr\}\,.
\ee

Considering the CHSH inequality as expressed by the quantity~\eqref{CHSHQuant} evaluated for the set of observables~\eqref{Observables} now computed in the transformed local frames, that is, for
\begin{equation}
\label{ObservablesPrime}
\begin{aligned}
\hat{A}_1^\prime =\cos{\Theta }\hat{S}_x^{(A)} - \sin{\Theta} \hat{S}_z^{(A)}&, \quad \hat{A}_2^\prime =\hat{S}_y^{(A)}, \\[2mm]
\hat{B}_1^\prime = -\frac{\cos{\Theta }\left( \hat{S}_x^{(B)} + \hat{S}_y^{(B)} \right) + \sin{\Theta} \hat{S}_z^{(B)} }{\sqrt{2}}&, \quad \hat{B}_2^\prime =\frac{ \cos{\Theta } \left( \hat{S}_x^{(B)} - \hat{S}_y^{(B)} \right)+ \sin{\Theta} \hat{S}_z^{(B)} }{\sqrt{2}},
\end{aligned}
\end{equation}
we finally get
\begin{equation}\label{coreq}
\mathcal{S}^\prime[\vert \psi^\prime \rangle] = 2 \sqrt{2} \cos^2 \Delta\,.
\end{equation}
This means that if a CHSH-like experiment is performed by only rotating the observables from~\eqref{Observables} to~\eqref{ObservablesPrime}, the degree to which the associated CHSH inequality is violated will be reduced by $\cos^2 \Delta$. In this sense, the quantity \eqref{coreq} carries a fingerprint of curved spacetime effects through the phase $\Delta$. It must be emphasized that the transformation is unitary, and thus the amount of quantum correlations encoded in the pure states $\vert \psi \rangle$ and $\vert \psi^\prime \rangle$ are the same, as well as the total amount of EPR correlations obtained by maximizing $\mathcal{S}$ over all possible sets of observables $\{ A_1, A_2, B_1, B_2 \}$. Such conclusions are in line with previous findings~\cite{ueda,spingrav}.

In the limit $\psi\to\phi\to\phi_{GR}$, where $\phi_{GR}=-GM/r$ is the gravitational potential of the Schwarzschild solution, we recover the weak-field behavior of Refs.~\cite{ueda}. However, in the general framework we are concerned with, such occurrence does not hold true, since $\phi\neq\psi$. Despite this, as already pointed out in different contexts~\cite{ours}, our current approximation allows us to cast the gravitational potentials as
\be\label{approx}
\phi=\phi_{GR}+\phi_Q\,, \qquad \qquad \psi=\psi_{GR}+\psi_Q=\phi_{GR}+\psi_Q\,,
\ee
in which the contribution due to GR and the one associated with extended models of gravity can be properly distinguished. Accordingly, Eq.~\eqref{delta} becomes
\be\label{delta2}
\Delta=\Delta_{GR}-\delta_Q\,,
\ee
where $\Delta_{GR}$ is the term associated with GR~\cite{ueda} which can be achieved by means of the substitution $\phi=\psi=\phi_{GR}$ in Eq.~\eqref{delta} giving
\be\label{deltag}
\Delta_{GR}=\varphi\Bigl[\cosh\zeta\lf(1+2\phi_{GR}\ri)-1\Bigr]\,.
\ee
On the other hand, 
\be\label{corr}
\delta_Q=r\varphi\cosh\zeta\,\pa_r\lf(\phi_Q+\psi_Q\ri)
\ee
is the correction due to quadratic theories of gravity.

The rotation induced by the term $\delta_Q$ is particularly interesting for the case in which $\phi_Q\neq\psi_Q$. In such a scenario, Eq.~\eqref{corr} can be split into two separate quantities, namely
\be\label{terms}
\delta_Q=2r\varphi\cosh\zeta\,\pa_r\phi_Q-r\varphi\cosh\zeta\,\pa_r\lf(\eta\,\phi\ri)=2r\varphi\cosh\zeta\,\pa_r\phi_Q-\delta_{SEP}\,,
\ee 
where 
\be\label{nord}
\eta=1-\frac{\psi}{\phi}\,,
\ee
is the so-called Nordtvedt parameter~\cite{nordtvedt} up to the weak-field limit and by assuming that all non-linear gravitational effects belong to the GR sector~\cite{ours}. As long as $\eta\neq0$, the strong equivalence principle (SEP) fails to hold~\cite{nordtvedt}. Therefore, this implies that SEP violation may entail a degradation of the EPR perfect anticorrelation. However, as it can be deduced from Eq.~\eqref{terms}, we observe that $\delta_{SEP}\neq0$ does not necessarily imply $\delta_Q\neq0$, since the contribution due to extended theories of gravity is not entirely related to SEP violation. As a matter of fact, in the following considerations we highlight how in some models, despite $\delta_{SEP}\neq0$, we get a vanishing correction ascribable to the quadratic models, and thus $\Delta=\Delta_{GR}$.

\section{Model examples}\label{VI}

In this Section, we evaluate the corrective term~\eqref{corr} due to several quadratic models of gravity. In so doing, we also pinpoint the explicit form that the SEP-violating phase $\delta_{SEP}$ assumes for each scenario.

\subsection{$\mathcal{R}^2$ gravity}

The most immediate generalization of GR predicts the existence of an additional $\mathcal{R}^2$ contribution in the Einstein-Hilbert action, which in our framework is realized in Eq.~\eqref{quad-action} with
\be\label{sel1}
\mathcal{F}_1=\al\,, \qquad \mathcal{F}_2=0\,,
\ee
where $\al$ is a constant form factor. Under such circumstance, the metric potentials become
\be\label{mp1}
\phi=\phi_{GR}\lf(1+\frac{1}{3}e^{-m_0r}\ri)\,, \qquad \psi=\phi_{GR}\lf(1-\frac{1}{3}e^{-m_0r}\ri)\,,
\ee
with $m_0=1/\sqrt{3\al}$.

We can observe that, in the lowest-order approximation we are currently dealing with, no further contribution to Eq.~\eqref{delta2} apart from $\Delta_{GR}$ arises at all. Indeed, despite the expression for $\delta_{SEP}$ yields
\be\label{sep1}
\delta_{SEP}=-\frac{2}{3}\,\phi_{GR}\,\varphi\,e^{-m_0r}\lf(1+m_0r\ri)\cosh\zeta\,,
\ee
such quantity is precisely balanced by the first term in the r.h.s. of Eq.~\eqref{terms}, thus leaving no further signature except for the standard GR one.

\subsection{Fourth-order gravity}

Stelle's fourth-order gravity~\cite{stelle} is based on the same picture explained in the previous example together with the introduction of another quadratic term built upon the contraction of two Ricci tensors, thereby modifying our starting action with the following functions of the d'Alembert operator:
\be\label{sel2}
\mathcal{F}_1=\al\,, \qquad \mathcal{F}_2=\beta\,,
\ee
where $\al$ and $\beta$ are the constant form factors of the theory. In a similar framework, the metric potentials are given by
\be\label{mp2}
\phi=\phi_{GR}\lf(1+\frac{1}{3}e^{-m_0r}-\frac{4}{3}e^{-m_2r}\ri)\,, \qquad \psi=\phi_{GR}\lf(1-\frac{1}{3}e^{-m_0r}-\frac{2}{3}e^{-m_2r}\ri)\,,
\ee
with $m_0=2/\sqrt{12\al+\beta}$ and $m_2=2/\sqrt{-\beta}$.

\begin{figure}[ht!]
  \centering
    \includegraphics[width=8.5cm]{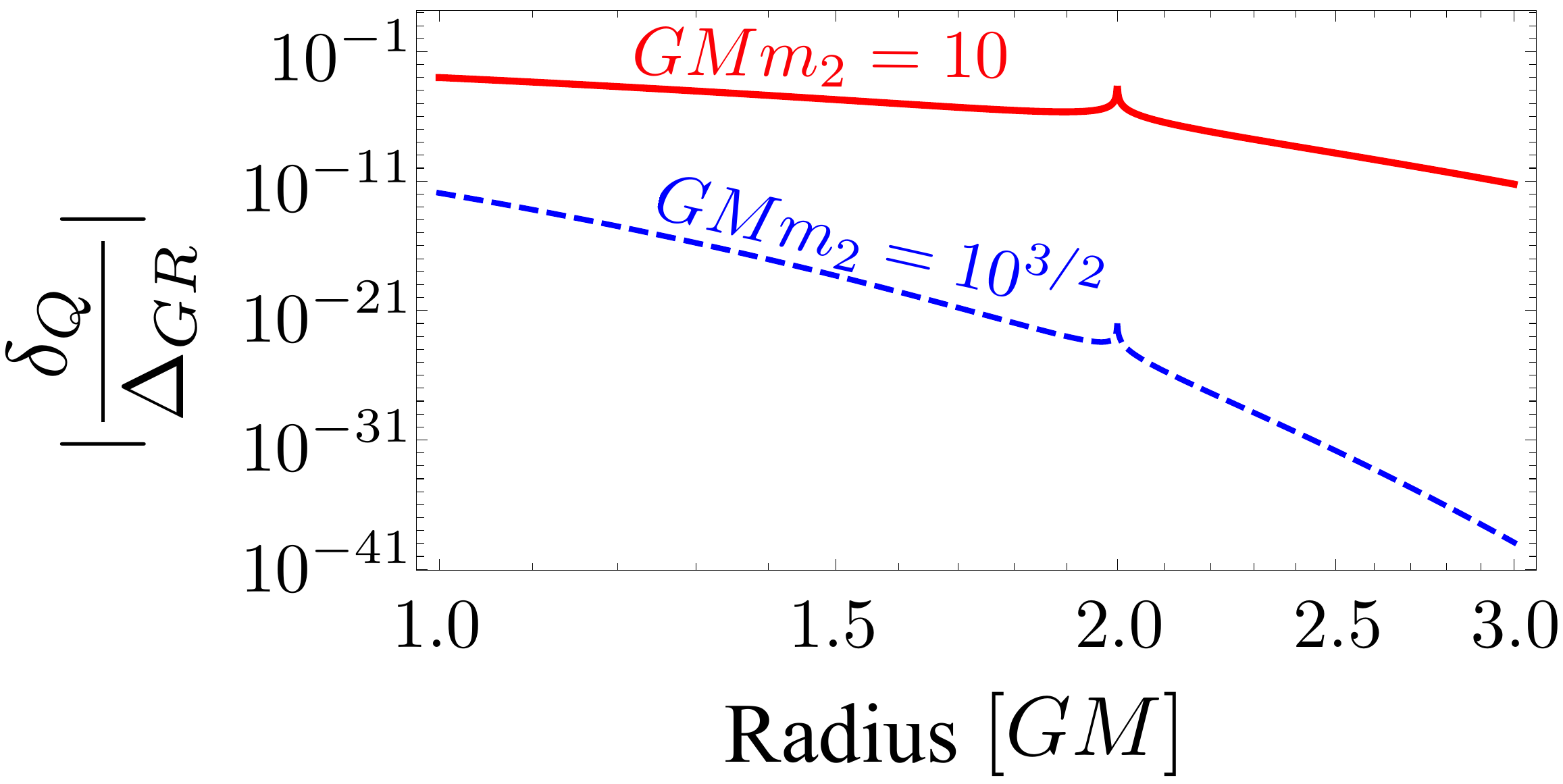}
    \caption{Ratio between the corrections $\delta_Q$ of the fourth-order gravity model and the GR corrections $\Delta_{GR}$ as a function of the dimensionless radius $r/GM$. The peak occurs at $r = 2GM$ (in natural units), which corresponds to the minimum point of $\Delta_{GR}$ (see Eq.~\eqref{deltag}). We have also adopted $\zeta = 10$.}
  \label{Plot1}
\end{figure}

Here, both $\delta_Q$ and $\delta_{SEP}$ are non-vanishing, and their respective expressions are 
\bea\non
\delta_Q&=&{2}\,\phi_{GR}\,\varphi\,e^{-m_2r}\lf(1+m_2r\ri)\cosh\zeta\,,\\[2mm]\label{res2}
\delta_{SEP}&=&-\frac{2}{3}\,\phi_{GR}\,\varphi\Bigl[e^{-m_0r}\lf(1+m_0r\ri)-e^{-m_2r}\lf(1+m_2r\ri)\Bigr]\cosh\zeta\,.
\eea
In Fig.~\ref{Plot1}, we depict $\vert \delta_Q /\Delta_{GR} \vert$ as a function of the radius in units of $GM$ and two values of the free parameter of the theory $m_2$. 

\subsection{Ghost-free infinite derivative gravity}

For the next example, we focus on a ghost-free nonlocal theory of gravity, which has been extensively studied in recent years~\cite{Biswas:2011ar,Biswas:2013cha,Biswas:2016etb,modesto,papers}. To keep our considerations simple, we deal with the easiest choice for the $\mathcal{F}(\Box)$ operators, which is represented by
\be\label{sel3}
\mathcal{F}_1=-\frac{1}{2}\mathcal{F}_2=\frac{1-e^{-\Box/M_s^2}}{2\Box}\,,
\ee
with $M_s$ being the scale at which nonlocal effects of gravity are relevant. In light of the above choice for the differential operators, the ensuing metric potentials coincide and their form is 
\be\label{mp3}
\phi=\psi=\phi_{GR}\,\mathrm{Erf}\lf(\frac{M_s r}{2}\ri)\,,
\ee
where
\be\label{erf}
\mathrm{Erf}(x)=\frac{2}{\sqrt{\pi}}\int_0^xe^{-z^2}dz
\ee
denotes the error function.

\begin{figure}[ht!]
  \centering
    \includegraphics[width=8.5cm]{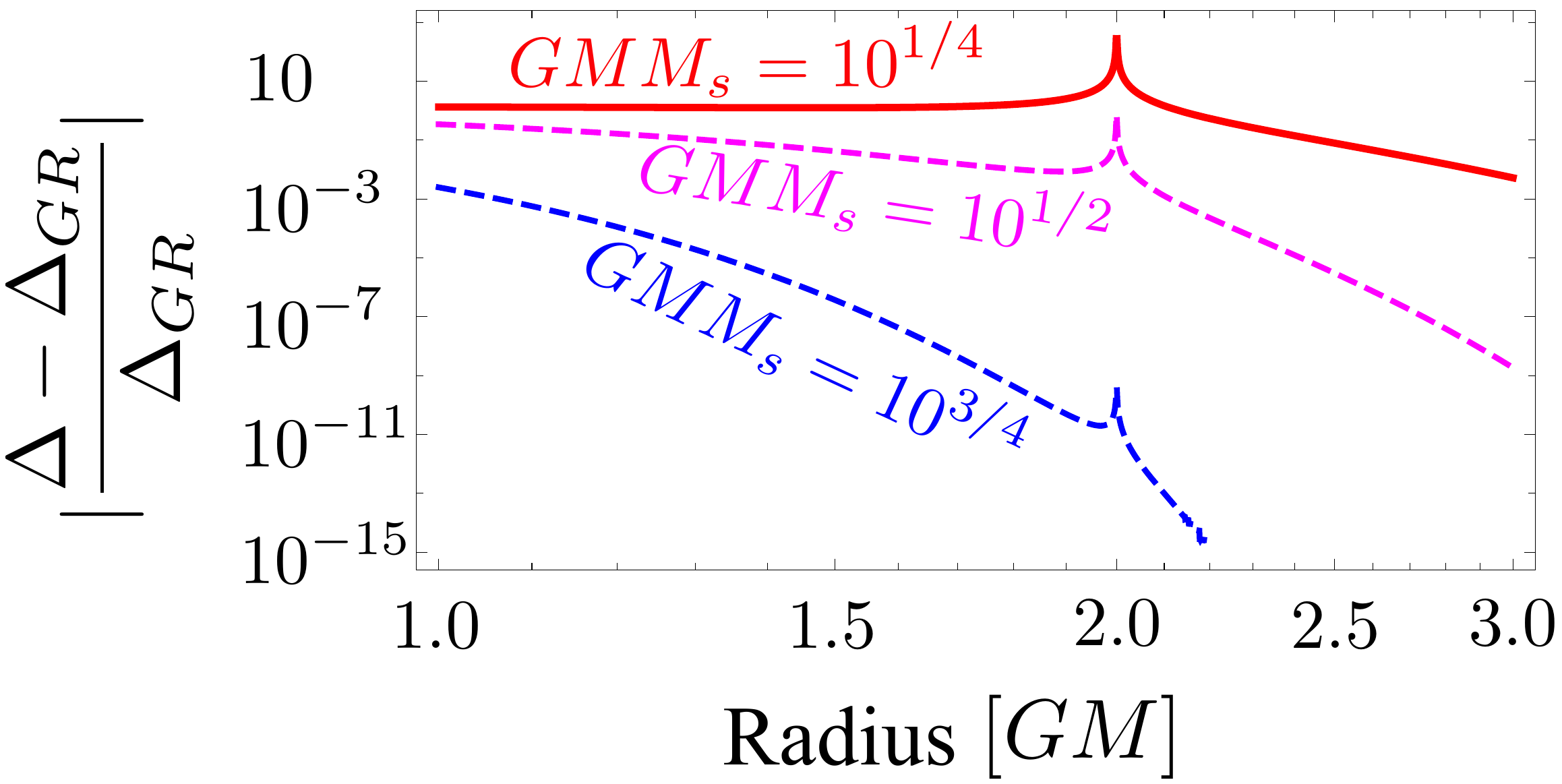}
    \caption{Ratio between the correction $\Delta - \Delta_{GR}$ of the ghost-free infinite derivative gravity and the GR correction $\Delta_{GR}$ as a function of the dimensionless radius $r/GM$. As in the case of Fig.~\ref{Plot1}, the peak occurs at $r = 2GM$ corresponding to the minimum point of $\Delta_{GR}$ (see Eq.~\eqref{deltag}). As before, we have adopted $\zeta = 10$.}
  \label{Plot2}
\end{figure}

Since the metric potentials are equal, we immediately deduce that $\delta_{SEP}=0$. However, the overall correction to the GR scenario $\delta_Q$ is different from zero, but because of Eq.~\eqref{mp3}, we cannot split $\Delta$ into $\Delta_{GR}$ and $\delta_Q$ analytically. The complete outcome which includes both GR and its extension reads
\be\label{res3}
\Delta=\varphi\Bigl\{\cosh\zeta\Bigl[1+2\phi_{GR}\,\mathrm{Erf}\lf(\frac{M_sr}{2}\ri)+2\,\frac{GMM_s}{\sqrt{\pi}}\,e^{-M_s^2r^2/4}\Bigr]-1\Bigr\}\,.
\ee
We plot $\vert (\Delta - \Delta_{GR})/\Delta_{GR} \vert$ as a function of the radius (in units of $GM$) in Fig.~\ref{Plot2} for three values of $M_S$. In passing, we notice that the corrections due to this extended model are less prominent than those induced by the fourth-order model (see Fig.~\ref{Plot1}), but at the same time they are also more sensitive to the variation of the free parameter $M_S$.

\subsection{Nonlocal gravity}

The last example we consider is an infrared extension of GR which draws inspiration from quantum correction to the effective action of quantum gravity~\cite{nonlocal}. Specifically, the simplest modification we can make to the Einstein-Hilbert action entails the following form of the differential operators appearing in Eq.~\eqref{quad-action}:
\be\label{sel4}
\mathcal{F}_1=\frac{\al}{\Box}\,, \qquad \mathcal{F}_2=0\,.
\ee
The previous equation implies a plain but essential change to the GR metric potentials, which in the present context are represented by
\be\label{mp4}
\phi=\phi_{GR}\lf(1+\frac{\al}{3\al-1}\ri)\,, \qquad \psi=\phi_{GR}\lf(1-\frac{\al}{3\al-1}\ri)\,.
\ee
As in the case of $\mathcal{R}^2$ gravity, the overall contribution due to the generalization of GR is vanishing (i.e. $\delta_Q=0$), even though the phase associated to the SEP violation is not, namely
\be\label{sep4}
\delta_{SEP}=-\frac{4\al}{3\al-1}\,\phi_{GR}\,\varphi\,\cosh\zeta\,.
\ee


A similar outcome occurs also for another nonlocal model of gravity~\cite{nonlocal,nl2} which suggests that the action shall be modified as follows
\be\label{sel5}
\mathcal{F}_1=\frac{\beta}{\Box^2}\,, \qquad \mathcal{F}_2=0\,.
\ee
By virtue of the above choice, the metric potentials become more elaborate, in that they acquire a further dependence on $r$, i.e.
\be\label{mp5}
\phi=\frac{4}{3}\,\phi_{GR}\lf(1-\frac{1}{4}\,e^{-\sqrt{3\beta}\,r}\ri)\,, \qquad \psi=\frac{2}{3}\,\phi_{GR}\lf(1+\frac{1}{2}\,e^{-\sqrt{3\beta}\,r}\ri)\,.
\ee
Also in this scenario, it is straightforward to observe that $\delta_Q=0$ whereas
\be\label{sep5}
\delta_{SEP}=-\frac{2}{3}\,\phi_{GR}\,\varphi\,e^{-\sqrt{3\beta}\,r}(1+\sqrt{3\beta}\,r)\cosh\zeta\,.
\ee


\section{Conclusions}\label{VII}

In summary, we have computed the effects of extended models of gravity on pure-state EPR correlations encoded in pairs of spin $1/2$ particles. As an EPR pair propagates in a circular trajectory, the degree of violation of the CHSH inequality with respect to flat space is reduced by a factor $\cos^2\Delta$ and we have shown that, in the linearized regime, the contributions of the extended models arising from corrections to the Einstein-Hilbert action can be readily identified and isolated. While for some extended models (i.e. $\mathcal{R}^2$ gravity and nonlocal gravity) such a contribution vanishes, in other cases it is nonzero and clearly decouples from the well-known general relativity effects~\cite{ueda}. It is worth pointing out that the factors that can be ascribed to quadratic models also encompass signatures of a violation of the strong equivalence principle, a feature already verified in other physical scenarios~\cite{ours}.

Our results can be used as a basis to design experiments aimed at probing gravitational effects beyond GR, although the setup would be impractical with massive particles. Nevertheless, the theory presented here can be generalized to a quantum optics framework in which the EPR source would produce a pair of photons with entangled polarizations whose correlations can be tested via interferometric techniques. A degradation of correlations beyond the one expected from GR would be a fingerprint of new physics phenomenology, and our calculations can be regarded as the starting point for obtaining bounds on the free parameters of the theory. Such experiments would require a sufficiently high number of photon pairs to achieve levels of sensitivity sufficient to resolve the small difference angle $\Delta - \Delta_{GR}$ (see Eq.~\eqref{delta2}). In fact, such a difference exclusively depends on the free parameters of the theory, but there are ``sweet spots'' where the effects due to the extended models are maximized, as we have shown in Figs.~\ref{Plot1} and \ref{Plot2}. These and other further possible developments are objects of current active investigation.

\section*{Acknowledgement}

FI and LP acknowledge support by MUR (Ministero dell’Universit\`a e della Ricerca) via the project PRIN 2017 ``Taming complexity via QUantum Strategies: a Hybrid Integrated Photonic approach'' (QUSHIP) Id. 2017SRNBRK. VASVB acknowledges financial support from the Max Planck Gesellschaft.

\end{document}